\documentclass[12pt]{article}

\title{PQ-Calculus of Fibonacci Divisors and Method of Images in Planar Hydrodynamics}

\author{Oktay K Pashaev \\
Department of Mathematics \\ Izmir Institute of Technology \\ Izmir 35430, Turkey}

\begin{document}

\maketitle              

\begin{abstract}
By introducing
the hierarchy of Fibonacci divisors and corresponding quantum derivatives,
we develop the golden calculus, hierarchy
of golden binomials and related exponential functions, translation operator and
infinite hierarchy of Golden analytic functions. The hierarchy of Golden periodic functions, appearing in this calculus we relate with the method of images in planar hydrodynamics
for incompressible and irrotational flow in bounded domain. We show that the even hierarchy
of these functions determine the flow in the
annular domain, bounded by concentric circles with the ratio of radiuses in powers of the Golden ratio. As an
example, complex potential and velocity field for the set of point vortices with Golden proportion of images
are calculated explicitly.
\end{abstract}


\section{Golden Ratio and Inversion in Circle}
The usual definition of Golden proportion or the Golden ratio is related with division of interval $x+y$ in proportion
$$  \frac{x+y}{x} =   \frac{x}{y} \,\,\,\,  \Rightarrow   \,\,\,\,  \varphi^2 = \varphi +1   , $$
were $x/y = \varphi = \frac{1 + \sqrt{5}}{2}\approx 1.6$ - Golden Ratio. Here we propose 
new definition of Golden Ratio, connected with  reflection in circle with radius $R$. Let  $a$ and $b$ are symmetric points
with respect to the circle at distance $R$ between them, satisfying equations 
$$a\, b = R^2 , \,\,\,\,\,\,  b-a = R.             $$
Then, distances to these points from origin are in Golden proportion of $R$,
$$a = \frac{1}{\varphi} R,\,\,\,\, b = \varphi R.$$ 
As well known, symmetric points with respect to circle at origin in complex plain are  $z$ and $R^2/\bar z$. These points correspond to position of a vortex and its image in the circle, 
according to method of images in hydrodynamics. Then, due to above definition, if the distance between vortex and its image is $R$, 
then positions of vortex and the image are in Golden proportion. For unit circle with $R=1$, these positions are $z = \varphi e^{i\theta}$ and $z^* = \frac{1}{\varphi} e^{i\theta}$.

If method of images is applied to problem with two circles, then an infinite set of images arises \cite{R1}. These images can be counted by $q$-periodic functions \cite{R2} 
and two circle theorem \cite{R3}
in $q$-calculus with $q=r^2_2/r^2_1$. For annular domain with two concentric circles of radiuses $r_1$ and $r_2$, it can be reformulated in terms of PQ-calculus, with $P=r^2_1$ and $Q=r^2_2$.
Then, the PQ number in this calculus 
$$   [n]_{PQ}   = \frac{P^n - Q^n}{P-Q}              $$
for $P=\varphi^k$ and $Q={\varphi'}^k$ becomes Binet formula for Fibonacci divisors (\ref{Binet}). This implies that calculus of Fibonacci divisors \cite{R4} can be applied to  problem of hydrodynamic images
in annular domain with two circles and the Golden ratio of images.

\section{Calculus of Fibonacci divisors} 
 The ratio of two Fibonacci numbers $F_n/F_m$ is not in general integer number. However, surprising fact is that $F_{k n}$, where $k, n \in Z$ 
is dividable by $F_k$. The infinite sequence of integer numbers  $$\frac{F_{k n}}{F_k} \equiv F^{(k)}_n$$ we call
Fibonacci divisors conjugate to $F_k$.   The Binet formula for these numbers
\begin{equation}
F^{(k)}_n = \frac{(\varphi^k)^n - ({\varphi'}^k)^n}{\varphi^k - {\varphi'}^k},\label{Binet}
\end{equation}
leads to recursion relation 
$$F_{n+1}^{(k)}=L_k F_{n}^{(k)}+(-1)^{k-1} F_{n-1}^{(k)},$$
  where $L_k$ are Lucas numbers.
  The first few  sequences of Fibonacci divisors $F_n^{(k)}$ for $k=1,2,3,4,5$ and $n=1,2,3,4,5,...$ are
\begin{eqnarray}
 k &=& 1 ;\,\,\, F_n^{(1)} = F_n = 1,1,2,3,5,...        \nonumber \\ 
 k& = &2 ;\,\,\, F_n^{(2)} = F_{2n} = 1,3,8,21,55,...    \nonumber\\ 
 k&= &3; \,\,\,F_n^{(3)}= \frac{1}{2} F_{3n}= 1,4,17,72, 305,...    \nonumber   \\  
 k&=&4;\,\,\, F_n^{(4)} = \frac{1}{3} F_{4n}= 1,7,48,329,2255,...    \nonumber \\    
 k&=&5;\,\,\, F_n^{(5)} = \frac{1}{5} F_{5n}=1,11,122,1353,15005,...   \nonumber
\end{eqnarray}

\subsubsection{Golden Derivatives} 

  The Golden derivative operator $_{(k)} D^{x}_{F}$ corresponding to Fibonacci divisors, conjugate to $F_k$, $k \in Z$
acts on arbitrary function $f(x)$ as
\begin{equation}
_{(k)} D^{x}_{F}[f(x)]=\frac{f(\varphi^{k} x)-f(\varphi'^{k} x)}{\left(\varphi^{k} -\varphi'^{k}\right)x}. 
\end{equation} 
For even $k$ in the limit $k \rightarrow 0$ it gives usual derivative
\begin{equation}
\lim_{k \rightarrow 0}  {_{(k)}} D^{x}_{F} f(x)  = f'(x)
\end{equation}
and
\begin{eqnarray}
_{(k)} D^{x}_{F} \, x^n = F_n^{(k)} x^{n-1}.  
\end{eqnarray}
The Leibnitz and the quotient rules for this derivative are
$$
_{(k)} D^{x}_{F}(f(x)g(x))= _{(k)} D^{x}_{F}(f(x))\phantom{.}g(\varphi^k x)+f\left(\varphi'^k x \right)\phantom{.}_{(k)} D^{x}_{F}(g(x)) ,
$$
$$
_{(k)} D^{x}_{F}\left(\frac{f(x)}{g(x)}\right)=\frac{_{(k)} D^{x}_{F}(f(x)) \phantom{.} g(\varphi^k x)-f(\varphi^k  x) \phantom{.} _{(k)}  D^{x}_{F}(g(x))}{g(\varphi^k  x)\phantom{.}g\left(\varphi'^k x\right)}.
$$

\subsubsection{Fibonacci Divisors and  Fibonomials}
	 The product of Fibonacci divisors,
\begin{equation}
F_1^{(k)} F_2^{(k)} \ldots F_n^{(k)}=\prod_{i=1}^{n} F_i^{(k)} \equiv F_n^{(k)}! \nonumber
\end{equation}
- the Fibonacci divisors factorial, can be considered as $k-th$ Fibonorial or generalized Fibonorial.
The  Fibonomial coefficients for Fibonacci divisors are
\begin{eqnarray}
_{(k)}  {n \brack m}_{F}=\frac{F_1^{(k)} F_2^{(k)} \ldots F_{n-m+1}^{(k)}}{F_1^{(k)} F_2^{(k)} \ldots F_m^{(k)}}=\frac{F_n^{(k)}!}{F_m^{(k)}! F_{n-m}^{(k)}!}.
\nonumber \end{eqnarray}
\subsubsection{Hierarchy of Golden Binomials}
The $k-th$ Golden binomial is defined by polynomial,
\begin{eqnarray}
_{(k)} \left(x-a\right)^{n}_{F} =\prod_{s=1}^{n}  \left( x-\varphi^{k(n-s)} \varphi'^{k(s-1)} a\right) \nonumber\end{eqnarray}
with following factorization rule
\begin{eqnarray}
_{(k)} \left(x-a\right)^{n+m}_{F}&=& _{(k)} \left(x-\varphi^{km} a\right)^{n}_{F} \phantom{.}  _{(k)} \left(x-\varphi'^{kn} a\right)^{m}_{F} \nonumber \\
&=& _{(k)} \left(x-\varphi'^{km} a\right)^{n}_{F} \phantom{.}  _{(k)} \left(x-\varphi^{kn} a\right)^{m}_{F}. \nonumber
\end{eqnarray}
It can be expanded in powers of $x$: 
\begin{eqnarray}
_{(k)} \left(x+y\right)^{n}_{F}=\sum_{m=0}^n    {_{(k)}} {n \brack m}_{F} (-1)^{k\frac{m(m-1)}{2}}x^{n-m} y^{m}. \nonumber
\end{eqnarray}
The $k-th$ Golden derivative acts on this binomial as 
\begin{eqnarray}
{_{(k)} D^{x}_{F}}\phantom{..} _{(k)} \left(x+y\right)^{n}_{F}&=& F_n^{(k)}\phantom{.}  _{(k)} \left(x+y\right)^{n-1}_{F},  \nonumber  \\
{_{(k)} D^{y}_{F}}\phantom{..} _{(k)} \left(x+y\right)^{n}_{F}&=& F_n^{(k)}\phantom{.}  _{(k)} \left(x+(-1)^k y\right)^{n-1}_{F}, \nonumber  \\
{_{(k)} D^{y}_{F}}\phantom{..} _{(k)} \left(x-y\right)^{n}_{F}&=&-F_n^{(k)}\phantom{.}  _{(k)} \left(x-(-1)^k y\right)^{n-1}_{F}. \nonumber
\end{eqnarray}
\subsubsection{Hierarchy of Golden functions}
 Let, entire complex valued function of complex variable z is
\begin{equation}
f(z)=\sum_{n=0}^\infty a_n \frac{z^n}{n!}. \nonumber
\end{equation}
 Then, for any integer $k$  exists entire complex function  
\begin{equation}
_{(k)} \phantom{.} f_{F}(z)=\sum_{n=0}^\infty a_n \frac{z^n}{F_n^{(k)}!}. \nonumber
\end{equation}
\subsubsection{Hierarchy of Golden exponential functions}
We introduce entire exponential functions 
\begin{eqnarray}
_{(k)} e^{x}_{F} &\equiv& \sum_{n=0}^\infty  \frac{x^n}{F_{n}^{(k)} !}, \nonumber\\
_{(k)} E^{x}_{F} &\equiv& \sum_{n=0}^\infty  (-1)^{k\frac{n(n-1)}{2}} \frac{x^n}{F_{n}^{(k)} !}. \nonumber
\end{eqnarray}
 The $k-{th}$ Golden derivative acts on these functions as
\begin{eqnarray}
_{(k)} D^{x}_{F} \left(_{(k)} e^{\lambda x}_{F}\right)&=&\lambda \phantom{..}   _{(k)} e^{\lambda x}_{F}, \nonumber \\
_{(k)} D^{x}_{F} \left(_{(k)} E^{\lambda x}_{F}\right)&=&\lambda \phantom{..} _{(k)}  E^{(-1)^k \lambda x}_{F}. \nonumber
\end{eqnarray}
Two exponential functions are related by formula
\begin{equation}
_{(k)}E^{x}_{F} = _{(-k)}e^{x}_{F}. \nonumber
\end{equation}
 The product of the exponentials is represented by series in powers of $k-$th Golden binomial
 \begin{equation}
_{(k)}E^{x}_{F} \cdot _{(k)}e^{y}_{F}=  \sum_{n=0}^\infty \frac{_{(k)}(x + y)^n_F}{F_n^{(k)}!}  \equiv _{(k)}e^{_{(k)}(x + y)_F}_{F}. \nonumber
\end{equation}
\subsubsection{Translation operator}
 $_{(k)}E^{y _{(k)}D^x_F}_{F}$ generates these binomials and $k-$th Golden functions
\begin{equation}
_{(k)}E^{y _{(k)}D^x_F}_{F} x^n = _{(k)}(x + y)^n_F, \nonumber
\end{equation}
\begin{equation}
_{(k)}E^{y _{(k)}D^x_F}_{F} f(x) = _{(k)}E^{y _{(k)}D^x_F}_{F} \sum_{n=0}^\infty a_n x^n = \sum_{n=0}^\infty a_n \cdot{_{(k)}(x + y)^n_F}. \nonumber
\end{equation}
\subsubsection{Hierarchy of Golden Analytic Functions}
By translation operator  we introduce complex valued $k-$th Golden analytic binomials
\begin{equation}
_{(k)}E^{iy _{(k)}D^x_F}_{F} x^n = _{(k)}(x + i y)^n_F \nonumber
\end{equation}
and the hierarchy of $k-$th Golden analytic functions
\begin{equation}
_{(k)}E^{i y _{(k)}D^x_F}_{F} f(x) =  \sum_{n=0}^\infty a_n \cdot{_{(k)}(x +i y)^n_F} \equiv f\left(_{(k)}(x + i y\right)_F),\nonumber
\end{equation}
for every integer $k$ satisfying the $\bar \partial$-equation
\begin{equation}
\frac{1}{2}\left(_{(k)}D^x_F + i _{(-k)}D^y_F\right) f\left(_{(k)}(x + i y\right)_F) = 0.\nonumber
\end{equation} 
For the real and imaginary parts of these functions
\begin{equation}
u(x,y) = _{(-k)} \cos_F\left( y _{(k)}D^x_F\right) f(x),\,\,\,\,\,v(x,y) = _{(-k)} \sin_F\left( y _{(k)}D^x_F\right) f(x),\nonumber
\end{equation}
we have the  Cauchy-Riemann equations
\begin{equation}
_{(k)}D^x_F u(x,y) = _{(-k)}D^y_F v(x,y),\,\,\,\,_{(-k)}D^y_F u(x,y) = - _{(k)}D^x_F v(x,y), \nonumber
\end{equation}
and functions are solutions of the hierarchy of Golden Laplace equations
\begin{equation}
\left(_{(k)}D^x_F\right)^2 \phi(x,y) + \left(_{(-k)}D^y_F\right)^2 \phi(x,y) = 0. \nonumber
\end{equation}
\subsubsection{Golden periodic functions}
 The set of Golden derivatives, determines hierarchy of Golden periodic functions for
every natural $k$. If function $f(x)$ is Golden periodic ($k=1$),
\begin{equation}
D^{x}_{F}(f(x))=0 \Longleftrightarrow f(\varphi x)=f(\varphi' x), \nonumber
\end{equation}
then, it is also periodic for arbitrary $k-th $ order Golden derivative,
\begin{eqnarray}
 D^{x}_{F}(f(x))=0 \,\,\,\,&\Rightarrow& \,\,\,\, _{(k)} D^{x}_{F}(f(x))=0, \nonumber\\
 f(\varphi x)=f(\varphi' x)\,\,\,\,&\Rightarrow& \,\,\,\, f(\varphi^k x)=f(\varphi'^k x), \nonumber
\end{eqnarray}
 for $k=2,3,\ldots$
But the opposite is not in general true. Indeed, function $$f(x)= \sin\left(\frac{\pi}{\ln \varphi^2} \ln |x|\right)$$ is Golden periodic function with $k=2$, but
it is not Golden periodic, since
\begin{equation}
D^{x}_{F}(f(x))=2 \phantom{.} \frac{\cos\left(\frac{\pi}{\ln (\varphi^2)} \ln |x| \right)}{(\varphi-\varphi')x}\neq 0 . \nonumber
\end{equation}
\section{Hydrodynamic Images and Golden Periodic Functions}
\subsection{Two dimensional stationary flow}
We consider incompressible and irrotational planar flow,  
\begin{eqnarray} div \,\vec{u} & =& 0 \Rightarrow
 u_1 = \frac{\partial\psi}{\partial y}, \,\,\,\,u_2 = -\frac{\partial\psi}{\partial x},\\
 rot\, \vec{u} & = & 0 \Rightarrow
  u_1 = \frac{\partial\varphi}{\partial x}, \,\,\,\,u_2 = \frac{\partial\varphi}{\partial y},\end{eqnarray}
where real functions $\varphi(x,y)$ and $\psi(x,y)$ are velocity potential and stream function, correspondingly. 
These functions are harmonically conjugate and satisfy Cauchy-Riemann equations,
$$\frac{\partial\varphi}{\partial x} = \frac{\partial\psi}{\partial y},\,\,\,
 \frac{\partial\varphi}{\partial y} = -\frac{\partial\psi}{\partial x}.$$
Combined together, they determine 
complex potential  $f(z) = \varphi + i \psi$, as analytic function $\frac{\partial}{\partial \bar z} f(z) = 0$, of 
$z = x + iy $. Corresponding
complex velocity $\bar V (z) =  \frac{\partial}{\partial z} f(z)$ is anti-analytic function of $z$.

For hydrodynamic flow in bounded domain, the problem is 
for given ${C}$-the boundary curve,
find analytic function (complex potential) $F(z)$, with
boundary condition
$$\Im F|_{ C} =  \psi|_{ C} = 0.$$
This equation determines the stream lines of the flow, such that
normal velocity to the curve $v_n |_{ C} = 0$. 
 \subsubsection{Two Circle Theorem} 
Applying two circles theorem \cite{R3} for flow $f(z)$, restricted to annular domain: ${1}<|z|<{\sqrt{\varphi}}$ between two concentric circles ${C_1}: |z| = 1$, ${C_2}: |z| = \sqrt{\varphi}$,
we get complex potential
$$F_{\varphi}(z) = f_{ \varphi}(z) + \bar f_{ \varphi} \left(\frac{1}{z}\right),$$
where ${ q} = \frac{r^2_2}{r^2_1} = {\varphi}$, { flow in even annulus} -

$$ f_{ \varphi}(z) = \sum^\infty_{n=-\infty} f({ \varphi}^n z),$$
and { flow in odd annulus} -
$$\bar f_{ \varphi} \left(\frac{1}{z}\right) = \sum^\infty_{n=-\infty} \bar f \left({\varphi}^n \frac{1}{z}\right).$$
\subsubsection{Golden ${\varphi}$-periodicity of flow}
The Golden periodicity
$$f_{\varphi}({\varphi} z) = f_{ \varphi}(z) 
\Rightarrow F_{ \varphi}({\varphi} z) = F_{ \varphi}(z)$$
implies that complex potential of the flow is invariant under Golden Ratio rescaling and as follows 
it is Golden ${ \varphi}$-periodic function,
$$D_z f_{ \varphi}(z) = \frac{f(\varphi z) - f(z)}{(\varphi -1)z}  = 0.$$
Corresponding complex velocity
 $$ \bar V(z) = \sum^\infty_{n=-\infty} {\varphi}^n \bar v ({\varphi}^n z) - \frac{1}{z^2}\sum^\infty_{n=-\infty} { \varphi}^n  
v \left({ \varphi}^n \frac{1}{z}\right)$$
is Golden $\varphi$-scale invariant function
$$ \bar V({\varphi} z) = {\varphi}^{-1} \bar V(z). $$
\subsubsection{Golden ${\varphi}$ scale-invariant analytic fractal} 
 For scale invariant function $f({\varphi} z) = {\varphi}^d f(z)$     $\rightarrow $  
  $$D_z f(z) = \frac{f({\varphi} z) - f(z)}{({\varphi}-1) z} = \frac{({\varphi}^d - 1)}{({ \varphi}-1) z} f(z),$$
and the ${\varphi}$-difference equation is
 $$ z D_z f(z) = [d]_{ \varphi} f(z).$$
Solution of this equation can be represented as 
$$f(z) = z^d A_\varphi (z),$$ 
where $$A_{\varphi} ({\varphi} z) = A_{\varphi}(z) $$ is arbitrary
 { $\varphi$}-periodic function, playing the role of 
$\varphi$-periodic modulation.
\subsubsection{Golden Weierstrass-Mandelbrot fractal}
As an example we consider
$$ W(t) = \sum^\infty_{n=-\infty} \frac{1 - \cos { \varphi}^n t}{{\varphi}^{n d}},\,\,\,\,\,0<d<1, \,{\varphi}>1,$$
- continuous but nowhere differentiable function, representing fractal with dimension $2-d$. It is
Golden self-similar function
$$ W({ \varphi} t )= {\varphi}^d W(t),$$
satisfying ${\varphi}$-difference equation
$$ t D_t W(t) = [d]_{\varphi} W(t). $$
By decomposing it as
$$W(t) = t^d A_{\varphi}(t)$$
we extract the Golden   ${\varphi}$-scale periodic part 
$A_{\varphi}({ \varphi} t) = A_{\varphi}(t)$, where
$$ A_{\varphi}(t) = \sum^\infty_{n=-\infty}  \frac{1 - e^{i {\varphi}^n t}}{({\varphi}^{d})^n t^{d}},$$
or in terms of Fibonacci numbers
$$ A_{\varphi}(t) = \sum^\infty_{n=-\infty}  \frac{1 - \cos ( {\varphi }F_n + F_{n-1}) t - i \sin ( {\varphi }F_n + F_{n-1}) t}{({\varphi}^{d})^n t^{d}}.$$
\subsubsection{Elliptic Function Form}
Let complex potential is Golden periodic analytic function
$ F(\varphi z) = F(z)$.
The Golden ratio can be represented 
$$   \varphi = e^{2\pi i \frac{\omega'}{\omega}}    $$
by arbitrary real $\omega$ and pure imaginary
$ \omega' = - i \frac{\omega}{2 \pi} \ln \varphi $.
Function
$$F(z) \equiv \Phi\left(\frac{\omega}{i \pi} \ln z \right) = \Phi(u)$$ is double periodic function
$$  \Phi(u + 2 \omega')    =  \Phi(u), \,\,\,\,
\Phi(u + 2 \omega)    =  \Phi(u) .           $$
It is elliptic function on Golden torus, determined by its singular points.
\subsubsection{Golden $\varphi$ periodic flow} 
Simplest example of Golden $\varphi$ periodic function (as principal branch) is
$$  F(z)   = z^{\frac{2 \pi i}{\ln \varphi}} = e^{\frac{2 \pi i}{\ln \varphi} \ln z}     = F(\varphi z)          . $$
Rewritten in polar coordinates $z = r e^{i \theta}$,
$$    F(z) = e^{- \frac{2\pi}{\ln \varphi} \theta}  \left(  \cos (2 \pi \log_\varphi r)   + i  \sin (2 \pi \log_\varphi r)  \right)                 $$
it gives
stream function
$$ \psi(r,\theta) =  e^{- \frac{2\pi}{\ln \varphi} \theta}    \sin (2 \pi \log_\varphi r)                                        $$
and complex velocity 
$$      \bar V (z) = \frac{dF}{dz}  = \frac{2 \pi i}{\ln \varphi}  { \frac{1}{z} } 
{  e^{\frac{2 \pi i}{\ln \varphi} \log_\varphi z} }    = \frac{\Gamma}{2 \pi i}    { \frac{1}{z}}   A_{\varphi}  (z) .               $$
In the last form it represents Golden modulated point vortex at origin with strength
	$ \Gamma = - \frac{4 \pi^2}{\ln \varphi} $ and 
	stream lines $\psi|_C = 0$ at $\sin (2 \pi \log_\varphi r)=0$ or
	$ 2 \pi \log_\varphi r =\pi n $, $n = 0, \pm 1, \pm 2,...$. These lines represent an infinite set of concentric circles 
	with radiuses 
	$$ r_n = \varphi^{\frac{n}{2}} .$$ 
	The ratio of two successive radiuses is the Golden Ratio  
	$$  q = \frac{r^2_{n+1}}{r^2_n}= \varphi . $$
	For the flow in 
	Golden annulus
	$r_0 =1$ and $r_1 = \sqrt{\varphi}$ we have $D_\varphi F(z) =0$,  
	and in $k$-th Golden annulus
	$r_0 =1$ and $r_k = \varphi^{\frac{k}{2}}$ it gives $D_{\varphi^k} F(z) =0.$ 

 Superposition
	$$     F_k (z)  =  \sum^{+\infty}_{N = -\infty}  a_N z^{\frac{2 \pi i}{k \ln \varphi} N}                        $$
	describes flow in circular annulus with radius $r = 1$ and $R = \varphi^{\frac{k}{2}}$,
	so that
	$$ F_k (\varphi^k z) = F_k (z) ,$$
and the flow is $\varphi^k$- periodic. 
\subsection{Vortex in Golden annular domain}
For point vortex at position $z_0$ in Golden annular domain, $1 < |z_0| < \varphi^{\frac{k}{2}}$, by Two Circle Theorem
$$F_k(z) = \frac{\Gamma}{2\pi i}\sum^\infty_{n=-\infty} \ln \frac{z - z_0 {\varphi}^{k n}}{z - \frac{1}{\bar z_0} {\varphi}^{k n}}$$
and
 $$
\bar V(z)= \frac{\Gamma}{2\pi i}   \sum_{n= - \infty}^{\infty} \left[
\frac{1}{z-  z_0 {\varphi}^{k n}}- \frac{1}{z-  \frac{1}{\bar z_0}{\varphi}^{k n}}\right].$$
The flow is Golden $\varphi^k$ periodic
$$ F_k({\varphi^k} z) = F_k(z), $$ 
with self-similar complex velocity 
$$ \bar V_k({\varphi^k} z) = \frac{1}{\varphi^k} \bar V(z).$$ 
It represents
modulation of point vortex by Golden periodic function $$\bar V(z)= \frac{\Gamma}{2\pi i\, z} A_{k} (z).$$
\subsubsection{Golden Ratio of pole singularities}
Pole singularities are located at positions
$${ z_n = z_0 \varphi^{k n}}$$
and 
at symmetric points 
$$z^*_n = { \frac{1}{\bar z_0} \, \varphi^{k n} },$$
where ${ n} = 0, \pm 1, \pm 2,...\pm\infty$. The ratio of two image positions is power of Golden ratio
$$   \frac{|z_{n+1}|}{|z_n|}  = \varphi^k .                         $$
The distance between symmetric points is growing in geometric progression 
$$  \left| z_n -  z^*_n     \right|    =   \left| z_0 -  z^*_0     \right|     \left(\varphi^{k} \right)^n.           $$

\subsubsection{Hierarchy of Golden Logarithmic Functions}
The set of vortex images is determined completely by singularities of the {$\varphi$-Logarithmic} function,
$$Ln_\varphi(1- z)\equiv -\sum_{n=
1}^{\infty}\frac{z^n}{{ [}n { ]_{ \varphi}}}.$$
It converges for $|z| < \varphi$, where {$\varphi$} - number 
$$
{ [}n{ ]_{ \varphi}} \equiv 1 + \varphi + \varphi^2 + ...+ \varphi^{n-1} = \frac{\varphi^{n}-1}{\varphi-1}$$
expressed by Fibonacci numbers is
${ [}n{ ]_{ \varphi}} = (F_{n+1}-1)\varphi + F_n.
$
More general function,
{ $\varphi^k$}-logarithm ($0 < |z| < \varphi^k$):
$$Ln_{\varphi^k}(1 + z) =\sum_{n=
1}^{\infty}\frac{(-1)^{n-1}z^n}{{ [}n { ]_{ \varphi^k}}} = \frac{1}{\varphi^k} \sum_{n=
1}^{\infty} \frac{z}{\varphi^{k n} + z}$$
is expressible by Fibonacci divisors
$  \varphi^{k n}  = \varphi^k F^{(k)}_n + (-1)^{k+1}  F^{(k)}_{n-1}   $ and 
 $$ [n]_{ \varphi^k} =  \frac{\varphi^k F^{(k)}_n + (-1)^{k+1}  F^{(k)}_{n-1} -1}{\varphi^k -1} . $$
It has an infinite number of simple pole singularities at $z = - \varphi^{k n}  $.

The logarithm function is related to entire exponential functions 
$$e_\varphi (z) = \sum_{n=0}^{\infty}
\frac{z^n}{{ [}n { ]_\varphi}!},\,\,\,\,
E_\varphi(z) = \sum_{n=0}^{\infty}
\varphi^{n(n-1)/2}\frac{z^n}{{ [}n { ]_\varphi}!},$$
which by Euler identities for $\varphi$-binomial can be written as infinite product
$$ e_\varphi\left(z\right) = E_{\frac{1}{\varphi}} \left(z\right) =\prod^\infty_{n=0} \left(1 + \frac{z}{\varphi^{n+2}}\right).$$
Zeroes of $\varphi$ - exp function
$$\varphi {Ln_\varphi(1-\alpha z)} = z\frac{d}{dz} \ln e_\varphi(-\varphi\alpha z)   $$
contribute to  complex potential
$$F(z) = \sum_{s=1}^N { i \kappa_s \ln (z - z_s)} +
\sum_{s=1}^N
i \kappa_s \ln
\frac{e_\varphi\left(-\varphi\frac{z}{z_s}\right)e_\varphi\left(-\varphi\frac{z_s}{z}\right)}
{e_\varphi\left(-\varphi{z \bar z_s}\right)e_\varphi\left(-\frac{\varphi^2}{z \bar z_s}\right)}, $$
so that all images in the second sum are determined by { zeros} of these functions.
Then, complex velocity is expressible as
\begin{eqnarray}
\bar V(z)= \sum_{s=1}^N {\frac{i \kappa_s}{z-z_s}} + \nonumber \\
\frac{i \varphi}{z} \sum_{k=s}^N \kappa_s \left(  Ln_\varphi
\left(1-\frac{z}{z_s}\right)- Ln_\varphi \left(1- {z\bar
z_s}  \right)\nonumber  + Ln_\varphi \left(1-\frac{\varphi}{z \bar
z_s}\right)- Ln_\varphi \left(1- \frac{z_s}{z}
\right)\right) \nonumber\end{eqnarray}
\subsection{Hydrodynamic Images and $k$-th Golden Derivatives}
 For even $k = 2l$,  the Fibonacci divisor derivative is determined by finite difference
\begin{eqnarray}
z\, _{(k)} D^{z}_{F}[f(z)]= \frac{f(\varphi^{k} z)-f(\frac{1}{\varphi^{k}} z)}{\left(\varphi^{k} - \frac{1}{\varphi^{k}}\right)} ,\nonumber
\end{eqnarray}
vanishing  for Golden periodic function
\begin{equation}
\, _{(k)} D^{z}_{F} F(z) = 0 .\nonumber
\end{equation}
In annular domain, bounded by circles $1 < |z| < \varphi^{\frac{k}{2}}$ the flow is  $k$-th Golden periodic $F_{k}(\varphi^k z) =F_{k}(z)$,
so that
\begin{equation}
\, _{(k)} D^{z}_{F} F_{k}(z) = 0.\nonumber
\end{equation}
\subsection{Single Vortex Motion}
For single vortex motion, subject to equation
$$\dot z_0 = \varphi\frac{i\kappa}{\bar z_0}\left[Ln_\varphi \left(1 - |z_0|^2  \right) - Ln_\varphi
\left(1 - \frac{\varphi}{|z_0|^2}
\right)\right],$$
the solution is described by
uniform rotation 
$
z_0(t)= z_0(0)
e^{i \omega t}$, with angular velocity
$${ \omega} = \frac{\varphi \kappa}{|z_0|^2}\left(Ln_\varphi\left(1 - |z_0|^2\right) - Ln_\varphi\left(1 - \frac{\varphi}{|z_0|^2}\right)\right).$$
The vortex is stationary $\omega = 0$   at geometric mean distance $|z_0| = \varphi^{\frac{1}{4}}$ and
ratio of frequencies at boundary circles is the Golden ratio
$$ \frac{|\omega_1|}{|\omega_2|} = \varphi .                $$
\subsubsection{Semiclassical quantization of vortex motion}
The Bohr-Zommerfeld quantization of single vortex motion gives 
discrete spectrum
 $$E_n = \frac{\Gamma^2}{4\pi} \ln \left| e_\varphi \left( -\varphi\left(n+\frac{1}{2}\right)\right)   e_\varphi\left( \frac{-\varphi^2}{ (n+ \frac{1}{2})}\right) \right|.$$
This expression never vanishes, since zeros of exponential functions in r.h.s. should satisfy following equations, $n + \frac{1}{2} = \varphi^{k+1}$ or $n + \frac{1}{2} = \varphi^{-k}$.
But in both equations the l.h.s is rational number, while the r.h.s. is irrational.

\subsection{N vortex dynamics}
For N - point vortices with circulations $\Gamma_1$,...,$\Gamma_N$,
at positions $z_1,...,z_N$, equations of motion are
\begin{eqnarray}\dot {\bar{z}}_n = { \frac{1}{2\pi i}\sum_{j=1 (j \neq n)}^{N}
\frac{\Gamma_j}{z_n-z_j}} +  \frac{1}{2\pi
i}\sum_{j=1}^{N}\sum_{n = \pm 1}^{\pm \infty}
\frac{\Gamma_j}{z_n-z_j \varphi^n} \nonumber - \frac{1}{2\pi
i}\sum_{j=1}^{N}\sum_{n = -\infty}^{\infty}
\frac{\Gamma_j}{z_n-\frac{1}{\bar z_n}\varphi^n}.\nonumber \end{eqnarray}
This is Hamiltonian system with 
Hamiltonian function
\begin{eqnarray}
H = { -  \frac{1}{4\pi}\sum_{i,j =1 (i\neq j)}^N \Gamma_i
\Gamma_j \ln |z_i - z_j|}  - \frac{1}{4\pi}\sum_{i,j =1}^N \Gamma_i
\Gamma_j \ln
\left|\frac{e_\varphi\left(-\varphi\frac{z_i}{z_j}\right)e_\varphi\left(-\varphi\frac{z_j}{z_i}\right)}
{e_\varphi\left(-\varphi z_i \bar
z_j\right)e_\varphi\left(-\frac{\varphi^2}{z_i \bar
z_j}\right)} \right|,\nonumber \end{eqnarray}
where the second sum describes an infinite set of images with Golden proportion of positions. The Green function
of the problem 
$$ G_I = { -\frac{1}{2\pi} \ln |z - z_l|}
-\frac{1}{2\pi}
\ln
\left|\frac{e_\varphi\left(-\varphi\frac{z}{z_l}\right)e_\varphi\left(-\varphi\frac{z_l}{z}\right)}
{e_\varphi\left(- \varphi z \bar
z_l\right)e_\varphi\left(-\frac{\varphi^2}{z \bar
z_l}\right)}\right|   + \frac{1}{4\pi}   \ln \varphi      $$
satisfies following conditions:1. symmetry $G_I (z, z_l) = G_I (z_l, z)$; 2. boundary values,
$G_I (z, z_l)|_{C_2} = 0$ - at the outer circle,
$
G_I (z, z_l)|_{C_1} = \frac{1}{2\pi} \ln \left|\frac{\sqrt{\varphi}}{z_l}\right|$ - at the inner circle.

Exact solution for N identical vortices $\Gamma_l = \Gamma, l = 1,...,N$, located at the same distance
$1 < r < \sqrt{\varphi}$ is
$$z_l (t) = r e^{i \omega t + i \frac{2\pi}{N}l },$$
where rotation frequency
$$\omega = \frac{\Gamma}{2 \pi r^2 }\left(\frac{N-1}{2}  + \varphi\sum_{j=1}^N
\left[Ln_\varphi \left( 1 - \frac{\varphi}{r^2}\, e^{i \frac{2\pi}{N}j}\right) -
Ln_\varphi \left( 1 - r^2\, e^{-i \frac{2\pi}{N}j}\right) \right]\right).$$
At geometrical mean distance $r= \varphi^{1/4}$ the frequency is
$$ \omega = \frac{\Gamma (N-1)}{4 \pi \sqrt{\varphi}}. $$       

%
%

\end{document}